\documentclass[journal]{IEEEtran}
\usepackage{blindtext}
\usepackage{graphicx}
\usepackage{float}
\usepackage{url}
\usepackage{amsmath}
\usepackage{color}
\usepackage{authblk}
\usepackage{listings}
\usepackage{multirow}
\usepackage{pbox}
\usepackage{color}
\usepackage[table,usenames,dvipsnames]{xcolor}
\usepackage{amsmath}
\usepackage{soul}

\definecolor{lightgray}{gray}{0.9}

\begin{document}
%

\title{Crowdsourcing Predictors of\\ 
Residential Electric Energy Usage}

\author{
    Mark~D.~Wagy, 
    Josh~C.~Bongard,
    James~P.~Bagrow,
    Paul D.~H.~Hines,~\IEEEmembership{Senior Member,~IEEE}%
    \thanks{This work was funded through grants from 
    the Burlington Electric Department, 
    the US Department of Energy (Award \#DE-OE0000315), 
    and the US National Science Foundation 
    (Award \#'s DGE-1144388, ECCS-1254549 and IIS-1447634. Human subjects work done under IRB exemption CHRBS: B09-225.
    }
    \thanks{M.~Wagy is with the Dept. of Psychological and Brain Sciences, Dartmouth College, Hanover, NH, USA: e-mail: mark.wagy@dartmouth.edu}
    \thanks{J.~Bongard is with the Dept.~of Computer Science and the Complex Systems Center, University of Vermont, Burlington, VT, USA: e-mail:jbongard@uvm.edu}%
    \thanks{J.~Bagrow is with the Dept. of Mathematics and Statistics ~and the Complex Systems Center, University of Vermont, Burlington, VT, USA, e-mail: james.bagrow@uvm.edu}%
    \thanks{P.~Hines is with the Dept.~of Electrical and Biomedical Engineering and the Complex Systems Center, University of Vermont, Burlington, VT, USA, e-mail: paul.hines@uvm.edu}%
}

\maketitle

\begin{abstract}
Crowdsourcing has been successfully applied in many domains including astronomy, cryptography and biology. In order to test its potential for useful application in a Smart Grid context, this paper investigates the extent to which a crowd can contribute predictive hypotheses to a model of residential electric energy consumption. In this experiment, the crowd generated hypotheses about factors that make one home different from another in terms of monthly energy usage. To implement this concept, we deployed a web-based system within which 627 residential electricity customers posed 632 questions that they thought predictive of energy usage. While this occurred, the same group provided 110,573 answers to these questions as they accumulated. Thus users both suggested the hypotheses that drive a predictive model and provided the data upon which the model is built. We used the resulting question and answer data to build a predictive model of monthly electric energy consumption, using random forest regression. Because of the sparse nature of the answer data, careful statistical work was needed to ensure that these models are valid. The results indicate that the crowd can generate useful hypotheses, despite the sparse nature of the dataset.

\end{abstract}

\section{Introduction}
\label{sec:introduction}
With the rapid adoption of Advanced Metering Infrastructure (AMI) \cite{wenpeng2009advanced}, end-users have an increased ability to track their electricity consumption~\cite{karnouskos2007advanced, ayres2012evidence, mohassel2014survey}. As a result, consumers have access to  data that might help them to understand what factors drive residential energy usage.

However, prior work~\cite{Attari:2010} has shown that consumers often misunderstand the relative impact of various loads on electricity usage. 
Information feedback can help consumers to find ways to reduce their energy usage~\cite{Asensio:2015}. But there is a need for research into tools that assist customers to interpret their electricity data to make better decisions. We hypothesize that with the right tools consumers will be able to lend their intuitions to assist in the construction of useful predictive models. To assess this hypothesis there is a need to design and test tools and methodologies that leverage smart meter data to help consumers understand their electricity consumption data.



Traditionally utilities have modeled residential electricity consumption with expert-driven surveys~\cite{Sanquist:2012}, which can help in the development of future usage scenarios. These models can then be used to provide feedback to consumers and to design energy efficiency programs. An alternative to expert-driven processes is to provide end-users themselves (i.e., the crowd) with tools that enable them to discover useful patterns through a collaborative process. Only very preliminary work has investigated the potential of crowdsourcing for helping residential customers to provide intuition to modeling methods usage~\cite{costanza2012understanding,wijaya2014crowdsourcing}. Hence, it is not known whether consumers, who are not experts in energy efficiency modeling, can add value to expert knowledge regarding residential energy usage.

In this study, we adapt the crowdsourcing technique described in~\cite{bongard2013crowdsourcing,bevelander2014crowdsourcing}, in which crowd participants ask questions that they believe drive some behavioral outcome, to the particular problem of predicting and explaining electric energy usage. Specifically, we ask users to contribute questions that they believe are predictive of electric energy usage and then these same participants are given the opportunity to answer questions contributed by their peers. From these questions and answers we build predictive models that indicate which behaviors are relevant in modeling energy usage. Based on these models, we ask: can a crowd of non-experts contribute to the process of hypothesis formulation about energy usage behaviors?

Prior research shows that expert insight into energy usage is a valuable contribution to energy modeling~\cite{Sanquist:2012}. 
From prior work we also know that insight from the crowd can be harnessed in the development of potentially interesting hypotheses~\cite{bongard2013crowdsourcing,bevelander2014crowdsourcing}. 
However, our prior application of this crowdsourcing technique to residential electricity usage modeling~\cite{bongard2013crowdsourcing} revealed that requiring customers to manually enter their electricity consumption was a substantial impediment to the implementation of this method. As a result the energy usage model in~\cite{bongard2013crowdsourcing} was inconclusive. 
Prior work~\cite{swain2015participation} has also shown that the data proceeding from this approach has a unique pattern of sparsity (many questions are left unanswered by many users), which is a challenge for generating useful predictive models from crowd-generated survey data.
Additional research is needed to understand how to use data generated by Advanced Metering Infrastructure systems, in combination with insight from the crowd, to generate models that can provide useful and adaptive feedback to people about their electricity usage.

In this study, in contrast to previous work~\cite{bongard2013crowdsourcing}, we are specifically interested to (a) determine whether the crowd is able to contribute predictive features \emph{in addition to those provided by experts} by actively linking together data from an AMI system and an online energy portal, and to 
(b) identify data modeling methods that can produce valid predictive models given the unique patterns of sparsity found in this type of crowdsourced data. 
The main contributions of this study are as follows. 
First, this paper presents a crowdsourcing method that can generate predictive models of electric energy usage from smart meter data, and thereby identify features that are important to energy usage in a particular population.
Second, we observe that data generated from this method have unique patterns of sparsity that are difficult for many conventional approaches to predictive modeling. The results suggest that random forest regression is particularly effective in dealing with this sparsity.

\section{Previous Work}
\label{sec:previous_work}
Crowdsourcing has been demonstrated to be effective at using human feedback to complete tasks that are presently difficult for computers by leveraging the pattern recognition abilities of humans~\cite{khatib2011algorithm}. Many examples of crowdsourcing have shown that humans can contribute valuable information to tasks that are considered difficult for machine learning methods. Studies have demonstrated that a loosely organized cohort of anonymous individuals are able to contribute to galaxy classification~\cite{lintott2008galaxy}, minimizing the energy configurations for protein folding~\cite{khatib2011algorithm} and rapid search in geolocation activities~\cite{pickard2011time}. A crowd has been shown to be effective at digitizing print material via web security enforcement puzzles~\cite{von2008recaptcha}, crowdsourced editing in a collaborative word processor~\cite{bernstein2010soylent} and even in building a massive online encyclopedia, as the web-based encyclopedia Wikipedia is itself a crowdsourced effort \cite{giles2005internet}.

That humans possess skills which, as of now, are not yet fully developed in modern machine intelligence methods suggests that there may be potential for humans to complement the capabilities of machine intelligence as part of human-machine collaborative systems. It remains to be seen what roles humans could most effectively contribute to such a hybrid intelligent system. One role that humans may effectively play is that of identifying hypotheses that can then be used in models generated by machine learning methods. Early thought in human-machine collaborative systems work theorized that humans could suggest ideas or hypotheses for which a machine counterpart could perform calculation and extrapolation tasks. The machine contribution would thus validate or disprove hypotheses developed by the human counterpart~\cite{licklider1960man}. More recently, it has been suggested that machines can be used to tie together disparate hypotheses~\cite{evans2010philosophy}. Initial investigations into the ability of the crowd in a hypothesis-generating role to determine what types of behaviors are predictive of body-mass-index and electricity consumption have shown promise~\cite{bongard2013crowdsourcing}.

Crowdsourcing has been applied to the solution of only a few electric energy problems. In~\cite{bongard2013crowdsourcing}, the authors describe preliminary findings for using a crowd feedback system to discover behavioral drivers for electricity usage. However, in this work, only in-sample modeling error was reported and there was no indication of whether crowd feedback was meaningfully incorporated into the models that resulted. In another recent project~\cite{wijaya2014crowdsourcing}, crowd participants were asked directly how consumers could be incentivized to reduce power consumption during peak usage hours. Other projects used monetary incentives to crowdsource energy forecasting \cite{humphreys2016crowdsourced} in contrast to the present work, which did not provide any monetary compensation to crowd participants. Some studies have employed crowdsourcing as solely a means for data-collection: collecting fine-grained information about building and home power usage~\cite{sagl2014crowdsourcing}, enabling smart grid in the absence of smart meters~\cite{karnouskos2011crowdsourcing} and using phones to identify power outages leading to blackouts~\cite{klugman2014grid}.

\section{Methods}
\label{sec:methods}
In the present study, we used the crowdsourcing concept introduced in~\cite{bongard2013crowdsourcing} to identify predictors of electric energy usage. This method proceeds as follows: First, participants are recruited to visit a website focused on understanding a behavioral outcome. Next, the participants are asked to answer a few questions. These are questions that others had previously suggested, which they believe to be effective predictors of the outcome of interest. For example, one might believe that obesity is related to eating habits and thus ask, ``How many meals do you eat per day?'' In the background a modeling engine works to identify relationships between the resulting answer data and the outcome of interest, and then communicates this information back to the participant.

In this paper we describe results from an application of this method to the problem of providing information feedback to residential electricity consumers. Specifically, our ``EnergyMinder'' application was designed to use AMI data from a small municipal utility, the Burlington Electric Department (BED), and the crowdsourcing method above to answer the following question: ``Why does one home consume more electric energy than another?''

This experiment began by designing a web site, which was made available to about 15,000 residential electric customers in Burlington, VT in the fall of 2013. With the exception of a small number of 'opt-out' households, all utility customers in this area have AMI or "Smart Meter" systems installed. After initially logging in to the site, customers could view their electricity consumption compared to that of others in the participant group, and then were invited to both answer and pose questions regarding residential power usage in the manner previously described. Once a user posed a new question, that question was forwarded to the moderators who verified that the question was not egregiously misleading and did not include personally identifying information, and then added to the crowdsourced survey. This process was seeded with a set of six expert-generated seed questions (see Table~\ref{tab:expert_questions}). Participants were free to answer and ask as many questions as they desired. Upon arriving at the ``ask'' page, the site prompted participants with the suggestion that they ask questions that they believed to be predictive of residential electricity usage. Users were not limited to answering or asking questions in a single session; they could return to the site as many times as they wanted to answer or ask questions.

The site allowed participants to pose three different types of questions: questions with numerical answers (e.g., ``How many loads of laundry do you do per week?''), yes/no questions (``Do you have access to natural gas?''), and agree/disagree questions (``I generally use air conditioners on hot summer days.''), which were based on a five-level Likert scale with the option to \textit{Strongly Disagree}, \textit{Disagree}, \textit{Neither agree nor disagree}, \textit{Agree} or \textit{Strongly Agree}. To initiate the question/answer process, a set of six seed questions, shown in Table ~\ref{tab:expert_questions}, were inserted into the EnergyMinder tool. 

\begin{table}[ht!]
\centering
\caption{Expert-generated seed questions}
\label{tab:expert_questions}
\begin{tabular}{l|p{3in}}
\textbf{ID} & \textbf{Text} \\ 
\hline
q1 & I generally use air conditioners on hot summer days \\    
q2 & Do you have a hot tub?                                \\   
q3 & How many teenagers are in your home?                    \\ 
q4 & How many loads of laundry do you do per week?             \\  
q5 & Do you have an electric hot water tank?                     \\
q6 & Most of my appliances (laundry machines, refrigerator, etc.) are high efficiency. 
\end{tabular}
\end{table}

\subsection{Online data modeling}
In addition, the site included a ``virtual energy audit'' page that was designed to provide some feedback to customers about factors that appeared to cause their energy consumption to deviate from the mean. To implement this, the site used a forward-only step-wise AIC (Akaike Information Criterion) linear regression approach to build a model of the form:
\begin{align}
 \Delta E_{30,i} = \sum_{j\in A} \beta_j z_{j,1} 
            + \epsilon_i
            \label{eq:reg}
\end{align}
where $\Delta E_{30,i}$ is the deviation of customer $i$'s energy consumption from the mean over the previous $30$ days, $A$ is the set of questions selected by the AIC algorithm, $\beta_j$ is the estimated coefficient for question $j$, $z_{i,j}$ is the mean (0) imputed z-score of user $i$'s answer to question $j$ after dropping outliers such that $|z_{i,j}|<3,\,\forall i,j$, and $\epsilon_i$ is the unexplained energy. $\Delta E_{30,i}$ was formed by summing the past four weeks of energy consumption data for user $i$, collected anonymously from the utility's AMI system, multiplying by $30/28$ to obtain a rolling one-month outcome value, and then subtracting from the mean consumption of this same period for the entire participant group.

To reduce the risk of over-fitting, the model was constrained to include no more than 20 terms. Given the model (\ref{eq:reg}), the energy audit page for participant $j$  displayed at most 10 questions for which the terms $|\beta_j z_{i,j}|$ were largest for this customer. This page also included an illustration of how much their answers to these questions impacted their predicted energy usage. In this way, users could see how their usage differed from usage patterns of an average consumer using questions and answers that had been found through the crowdsourcing process.

While this simple online model allowed the site to provide online feedback to participants, it also had a number of important limitations. First, without testing the model on out-of-sample data the predictive accuracy of the online model was not explicitly validated. Second, the sparse nature of the dataset (see Fig.~\ref{fig:na_image}), presented challenges for this type of mean-imputed linear model. We addressed these issues by exploring a variety of alternative modeling methods after the conclusion of the online research; see Sec.~\ref{sec:ex-post}.

\subsection{Differences from standard survey research}

\begin{figure}[H]
\centering
\includegraphics[width=1\columnwidth]{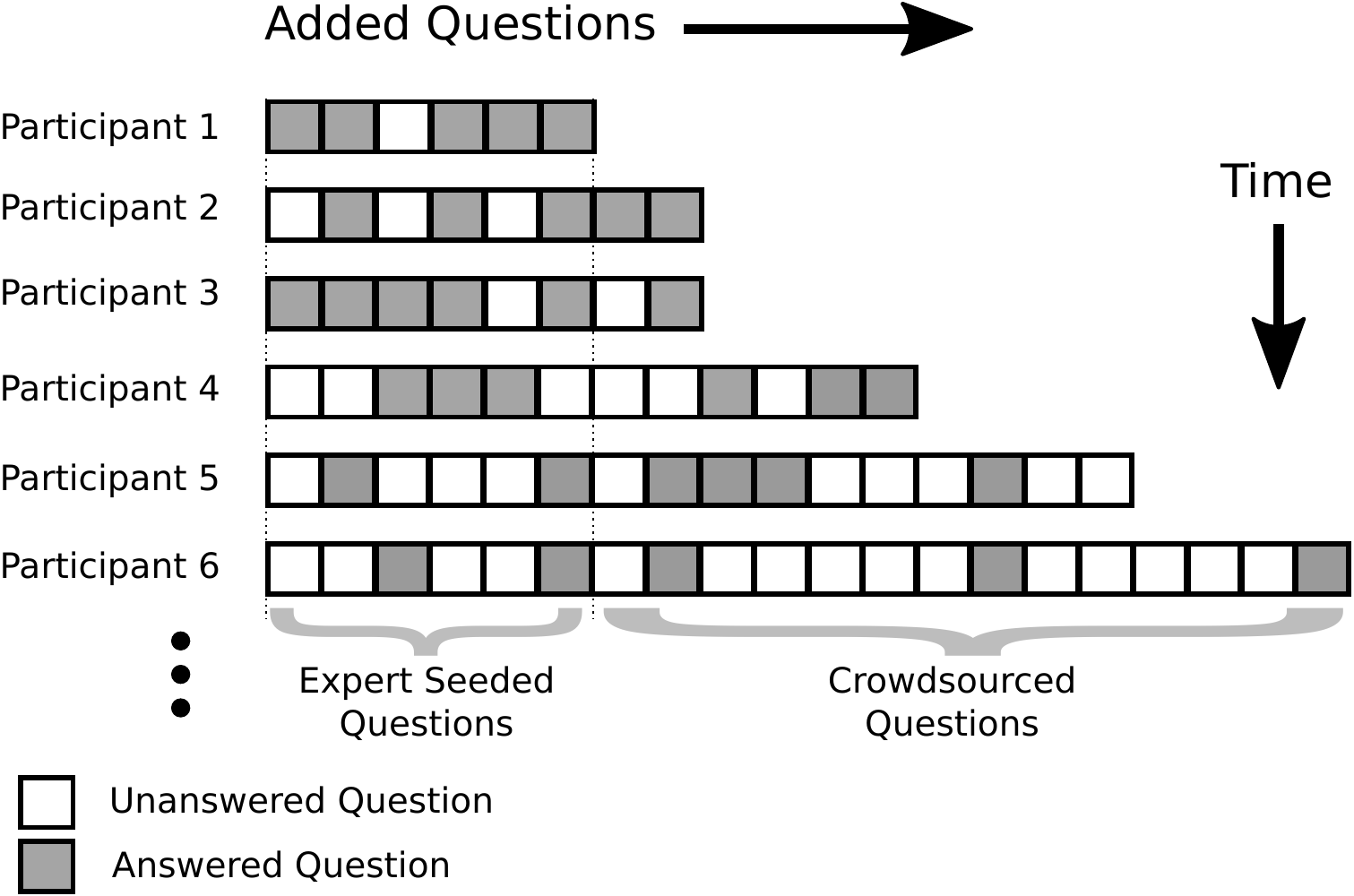}
\caption{Crowdsourcing questions and answers. The process begins with a set of seed questions for which the first participant contributes answers. The first participant then contributes two of their own questions. The second participant contributes answers to a sampling of the available questions, both seeded questions and questions contributed by the previous participant. This process continues over a set time period. As questions are added, the sparsity of the dataset grows because many of the questions contributed late in the process were not available to early participants. These questions go unanswered by earlier participants unless the participant returns to the site.}
\label{fig:questions_and_answers}
\end{figure}

This method of crowdsourcing survey information differs from standard survey-based research in that the participants are both generating the survey questions as well as answering them. When a participant poses a new question, they are effectively proposing a new crowd-generated hypothesis regarding behaviors that they believe may affect residential energy consumption by asking questions and provides data for predictive models by answering these questions. Thus a collaborative process exists in the way that the crowd participates: the number of questions are ever-growing as are the answers in response to that growing body of questions (see Fig.~\ref{fig:questions_and_answers}). In order to differentiate this type of crowdsourcing from the more common technique in which participants are asked to accomplish a fixed set of tasks, we call this process `collaborative crowdsourcing'.

Note that this type of question/answer feedback system necessarily results in a very sparse dataset. For example, consider the question posed by the last user of the system. This question would have only one answer associated with it since only the last user could answer it. Similarly, a question posed by the second-to-last user of the system would only not have more than two answers; and so on. What results from such a system is necessarily a sparse dataset of questions and answers, thus presenting a challenge for building a predictive model using only the crowd feedback.

\subsection{Ex post data analysis \label{sec:ex-post}}

After the experiment concluded, we revisited the resulting dataset in order to better understand whether the crowd could contribute to a predictive model of energy usage and to identify modeling methods that can most effectively identify valid patterns in the uniquely sparse datasets that result from this approach. 
In order to minimize spurious monthly variability, we built new outcome variables $E_{90,i}$ by summing the total electricity energy consumption for each customer $i$ from the utility-provided AMI data during a 90-day time period that had both high customer-participation in the EnergyMinder tool and when there was high (winter) electricity demand: from December 21, 2013 to March 21, 2014. 
This outcome was then used, along with the user-contributed questions and answers to build a predictive model and identify user-contributed questions (alone and in combinations) that were predictive of energy usage. The resulting dataset, $\mathcal{D}$, related each participant in the study (consisting solely of residential energy consumers), $i$, to a user-contributed question, $j$. The presence of a value at $\mathcal{D}_{i,j}$ indicated that user $i$ contributed an answer to question $j$. 

Several different model construction methods, including variants of the stepwise regression used in the online model, symbolic regression (genetic programming), LASSO, and decision trees, were tested to identify models that could reliably predict $E_{90,i}$.
After obtaining mixed results from these approaches, we focused on random forest regression  
\cite{breiman2001random,liaw2002classification}, which was able to reliably build predictive models from the uniquely sparse data that results from this type of crowdsourcing. Random forest regression is an ensemble-based machine learning method \cite{dietterich2000ensemble}, which consists of training a set of $k$ weak learner models $r(\textbf{x}_k,\Theta_k)$ on subsets of data as well as a subset of features using a process known as bagging. Bagging has been shown to be appropriate for building models on data that can, if perturbed, greatly alter the performance of the learned model and for 'data with many weak inputs' \cite{breiman1984classification}. An overall classification or regression prediction is then obtained by averaging the output of the weak learners to obtain a predictive model, $\mathcal{P}$. This process enables random forests to accommodate nonlinearities in the input data and without requiring assumptions of the underlying data distributions. The robustness to missing values and the ability to accommodate nonlinearities are the primary reasons that we chose random forest regression as our modeling method for the ex post analysis.

The explanatory features being used to build the model -- user-generated questions -- consisted of both numeric- and categorical-valued data. Due to the large degree of sparsity inherent in the process of collaborative crowdsourcing, our expectation was that the model fit would include a large amount of noise. However, we were mainly interested in whether some signal could be found in that noise, however slight. If $\mathcal{P}$ could outperform a model not incorporating user input; and it uses user-generated features despite the sparsity challenges, we could conclude that it had some value in explaining behaviors that contributed to residential energy usage. And it would thus demonstrate that the crowd indeed contributed to a predictive model.

The random forest regression algorithm requires that all values be present. To obtain a full dataset, we utilized the method of mean imputation. This imputation choice was governed by necessity: standard imputation methods, such as list-wise or pair-wise deletion would have resulted in a dramatic reduction in samples on which a predictive model could be built, rendering the modeling process impracticable. And due to the idiosyncrasies of this particular type of dataset, in which there is little overlap in answered values across features, methods that attempt to estimate joint distributions such as Bayesian multiple imputation methods~\cite{little2014statistical} were not able to converge (using standard settings for the \textit{mi} package \cite{gelman2011opening} of $30$ iterations and $4$ chains). After performing mean imputation, we normalized all values to z-scores.

To demonstrate that our model $\mathcal{P}$ could find some signal in the sparse dataset, $\mathcal{D}$, we compared it to a null model, $\mathcal{P}_\mathrm{null}$. This null model was trained using the same random forest regression algorithm. However, it was trained on a randomized (shuffled) version of $\mathcal{D}$ (denoted $\mathcal{D}^\mathrm{shuf}$). $\mathcal{D}^\mathrm{shuf}$ was obtained by randomly reordering user answers for each user-contributed question (along the margin, $\mathcal{D}^\mathrm{shuf}_{*,j}$). This had the effect of maintaining the basic statistical properties of $\mathcal{D}^{shuf}$ along the feature margins while disassociating the contributed answers with each user and their energy usage totals. If we were able to build predictive models $\mathcal{P}$ with consistently lower error than $\mathcal{P}_\mathrm{null}$, then we can conclude that the set of features used in the model are indeed predictive.

From random forests, we obtained a large set of decision trees whose leaves are aggregated to obtain a prediction for each set of inputs. The growth of such trees is governed by information reduction branching criteria on the available features. Despite the difficulty in interpreting a large collection of decision trees, which form a basis for random forests, we can obtain variable importance rankings \cite{breiman2001random}. We used the Gini impurity index to build a ranking of the features used in the model \cite{archer2008empirical}. Thus if crowd-generated questions have high rankings, we can reasonably assume that the crowd did indeed contribute useful information to the model.

However, we do not expect that all features in this ranking are contributing meaningfully to the predictive model. That is, there is a point at which the ranking of features transition from those contributing to the regression model and those that are included only by chance. To differentiate between the features that are meaningful versus those that are included only by chance, we utilized a technique for estimating the random degeneration between lists \cite{hall2012moderate}. In this method, we can obtain a cutoff point at which a set of ranked lists begin to diverge into random orderings \cite{schimek2015topklists}.

To obtain error estimates for $\mathcal{P}$ and $\mathcal{P}_\mathrm{null}$, we trained a set of $10$ independent Random Forest Regression models. We used the $10$ associated feature importance rankings for $\mathcal{P}$ for comparison. We compared these $10$ lists to estimate a value $k$ at which they began to deviate from a meaningful ranking of features that were consistently used in the model rather than being included only by chance. Those features that are ranked higher (i.e. lower in rank number) than this degeneration cutoff, $k$, are considered to be used in the predictive model not by chance, whereas those above $k$ were included due only to chance. As this method relies on parameters ($\nu$ and $\delta$, see \cite{hall2012moderate}), we performed a sensitivity study over a range of parameterizations.

\section{Results}
\label{sec:results}
From the period in which EnergyMinder was deployed from June 25, 2013 until September 24, 2014 a total of $627$ active crowd-participants (those who answered at least one question) asked $632$ questions, and provided $110,573$ answers to questions. 
These inputs were used to build predictive models for the mean monthly customer energy usage during the winter of 2013/2014, which was $514$ kWh/month with a standard deviation of $316$ kWh/month. This usage data approximately followed a log-normal distribution.

\begin{figure}[!ht]
\centering
\includegraphics[width=0.48\textwidth]{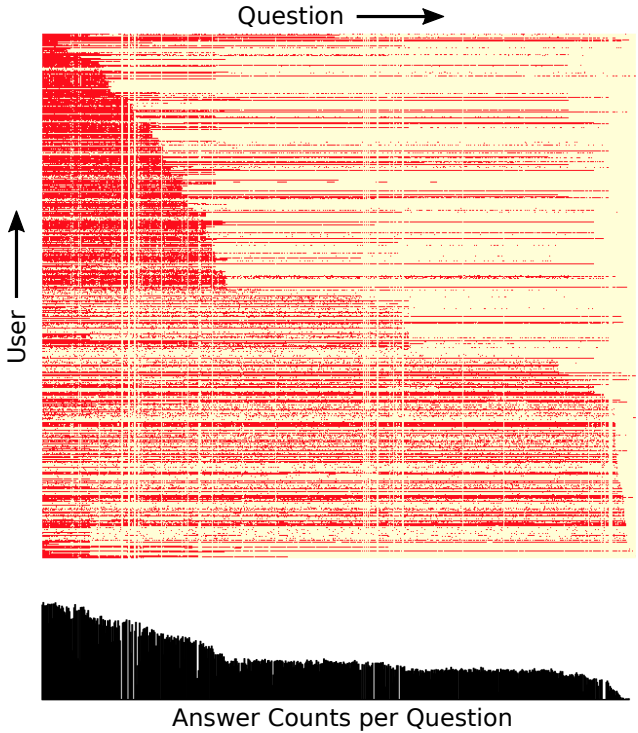}
\caption{Missing value pattern plot. Each red dot represents a user answering a question. Questions are shown along the horizontal dimension and users are on the vertical dimension. Total counts of answers per question are shown in the barplot in black. Note that users were given sequential id numbers such that customers that joined later received larger user id numbers.
}
\label{fig:na_image}
\end{figure}

Of the $632$ questions posed by the crowd, $627$ were answered at least once. 
Figure~\ref{fig:na_image} shows the pattern of missing values in the resulting dataset. 
In this plot users are ordered by the time that they signed up to participate in the study. The amount of sparsity in a question ranges from a maximum of $100\%$ missing to a minimum of $32.1\%$ missing.

Table~\ref{tab:notshuffled} shows representative top questions in ranked order by the random forest regression method described in Section~\ref{sec:methods}. The average mean squared error for the predictive model $\mathcal{P}$ 
was $0.883$ and the average mean squared error for $\mathcal{P_\mathrm{shuf}}$ was $1.031$, which was significant at $p<0.0001$ ($p=0.00001083$; Kolmogorov-Smirnov test, $D=1$), indicating a difference between the null model and the predictive model utilizing crowd-generated questions and data.


\begin{table*}[ht]
\centering
\caption{Top $43$ user-contributed questions as determined by random forest regression. Expert-created question IDs in bold.}
\begin{tabular}{ll}
\textbf{Question ID} & \textbf{Question Text} \\ \hline
\textbf{q4} &                                                              How many loads of laundry do you do per week? \\
q76       &                                                                      How many TVs are in your home? \\
\textbf{q1}&                                                         I generally use air conditioners on hot summer days \\
q24         &                      How many hours of TV or DVD/Video viewing is there, in your home, per week? \\
q142         &                                               Do you use washer and dryers outside of your home? \\
q13           &                                          How large is your home in square feet of living space? \\
q335           &                                             I know most of my neighbors on a first name basis. \\
q109           &                       Do you use heat tape during the winter on the pipes of your mobile home?\\
q34            &                                      How many months of the year do you use your dehumidifier?\\
q77            &                                      How many hours a day is there someone awake in your home?\\
q18            &                                                                    Do you have a dehumidifier?\\
q7             &                                                        How many people live in your household?\\
q12            &                                                          Do you have central air conditioning?\\
q22            &                                       How many pieces of toast do you toast on a typical week?\\
q62            &                            How high (in feet) are the ceilings on the main floor of your home?\\
q86            &                              Do you live in a rental unit that supplies your hot water heater?\\
q283           &                     I'd rather invest in energy efficient appliances than save for retirement.\\
q8             &                                                      I only run the dishwasher when it's full.\\
\textbf{q6}    &                       Most of my appliances (laundry machines, refrigerator, etc.) are high efficiency.\\
q56            &                  How many rooms in your home have an exterior wall and windows that face east?\\
q74            &                                     What's the size of your home or apartment, in square feet?\\
q167           &                                                    Do you use solar powered exterior lighting?\\
q81            &                                                    What temperature is your thermostat set at?\\
q54            &                                               Do you use a garbage-disposal unit in your sink?\\
q290           &                                 We should leave porchlights on as a courtesy to our neighbors.\\
q63  &How high (in feet) are the ceilings on upper/additional floors of your home? (Answer 0 if not applicable)\\
q107  &                                                                           Do you live in a mobile home?\\
q30   &                                             How many cars are generally parked at your home each night?\\
q72   &                                                            What percentage of your lights have dimmers?\\
q17   &                                                     Most of my lighting is high efficiency CFLs or LEDs\\
q95   &                                                             How long is your typical shower in minutes?\\
q2    &                                                                                  Do you have a hot tub?\\
q9    &                                                                     I only use lighting when necessary.\\
q121  &                                                          How many double paned windows are in your home\\
q11   &                                                 How many room/window air conditioners are in your home?\\
q141  &                                                  How many times a month do you eat away from your home?\\
q332  &                                      When presented with options for food sources, I usually buy local.\\
\textbf{q3}    &                                                                    How many teenagers are in your home?\\
q43            &                                     At what temperature is your electric hot water heater set?\\
q20            &                                                                  Do you have an electric oven?\\
q114           &                                           How many months per year do you hang clothes to dry?\\
q38            &                                                                  Do you have a microwave oven?\\
q25            &                                                              My desktop computer is always on.\\

\end{tabular}
\label{tab:notshuffled}
\end{table*}

The list-wise random deviation method for obtaining cutoffs in ranked lists (described in Section~\ref{sec:methods}) resulted in cutoff values $k$ that ranged between $9$ (for $\delta=1$ and $\nu=2$) and $89$ (for $\delta=10$ and $\nu=9$). Our sensitivity study over the parameters, $\delta$ and $\nu$ was run over a range of $[1,10]$ for $\delta$ and $[2,20]$ for $\nu$.

We also ran the random deviation cutoff method on $\mathcal{D}_{null}$ to obtain a list of ranked user-contributed questions. We did this for the same range of parameters as in the models trained on $\mathcal{D}$. Out of all possible parameterizations, $56\%$ instances of running the algorithm were not able to find a valid cutoff value that differentiated the meaningful rank comparisons from the ranked features that were included only by chance. Of the times that the algorithm was able to find a valid cutoff in the question ranking, $56$ cutoff values of $5$, $22$ cutoff values of $6$, and $6$ cutoff values of $7$.

\section{Discussion}
\label{sec:discussion}
\subsection{Contributed Questions and Participation}

\begin{figure}[!ht]
\centering
\includegraphics[width=9cm,height=8cm]{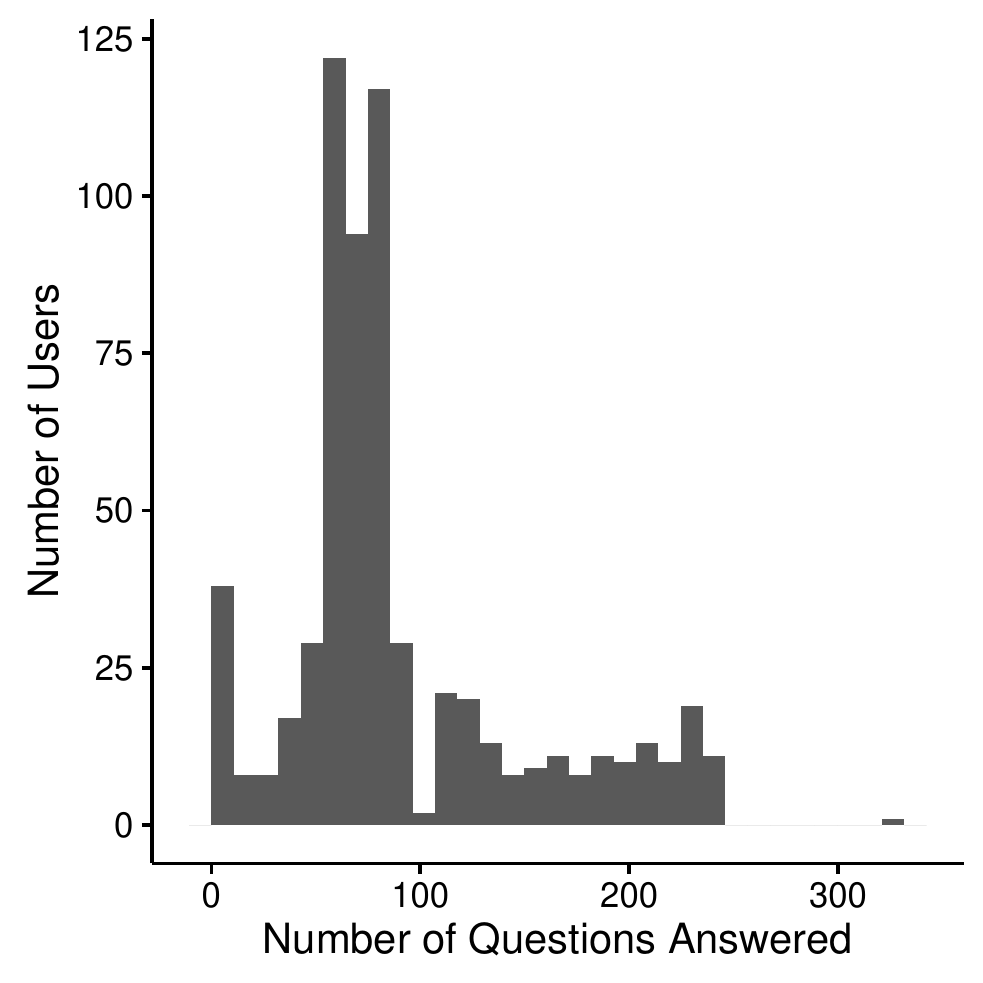}
\caption{Histogram showing user participation. A large number of users answered more than $100$ questions in total.}
\label{fig:answer_counts}
\end{figure}

The number of users providing answers ($627$) was very close to the number of questions in the system ($632$), thus creating a dataset that is approximately as `wide' as it is `long'. The median number of questions answered by a single user was $105$ questions and the maximum number of questions that a single user answered was $585$. A surprising number of users answered hundreds of questions (see Figure~\ref{fig:answer_counts}). A total of $166$ users answered more than $100$ questions. However, at least $32\%$ sparsity is present in all of the features used in each of the models. Thus the most completely populated feature contains answers from only $\sim68\%$ of users. And only a subset of these overlap with answers in other features. Starting at the top of Fig.~\ref{fig:na_image} and moving down to the bottom, we can see a widening band of answered questions starting as users increasingly participate. The questions outside of this band indicate that users did indeed return to the study to answer questions that were posed after they had visited the study for the first time. At roughly the vertical middle of the figure we see a point at which a large number of questions were added by either a single or just a few users as indicated by the rapid increase of number of questions answered. Future work could investigate whether providing incentives to return and answer questions to reduce the proportion of missing values in the data.

There are prominent vertical bands in Figure~\ref{fig:na_image} indicating questions without any answers (or very few answers). While users were given the opportunity to skip a question, most of these empty questions can be explained by those questions that were rejected by the moderator for not following directions, being offensive or asking respondents to reveal too much personal information.

\begin{figure}[!ht]
\centering
\includegraphics[width=9cm,height=8cm]{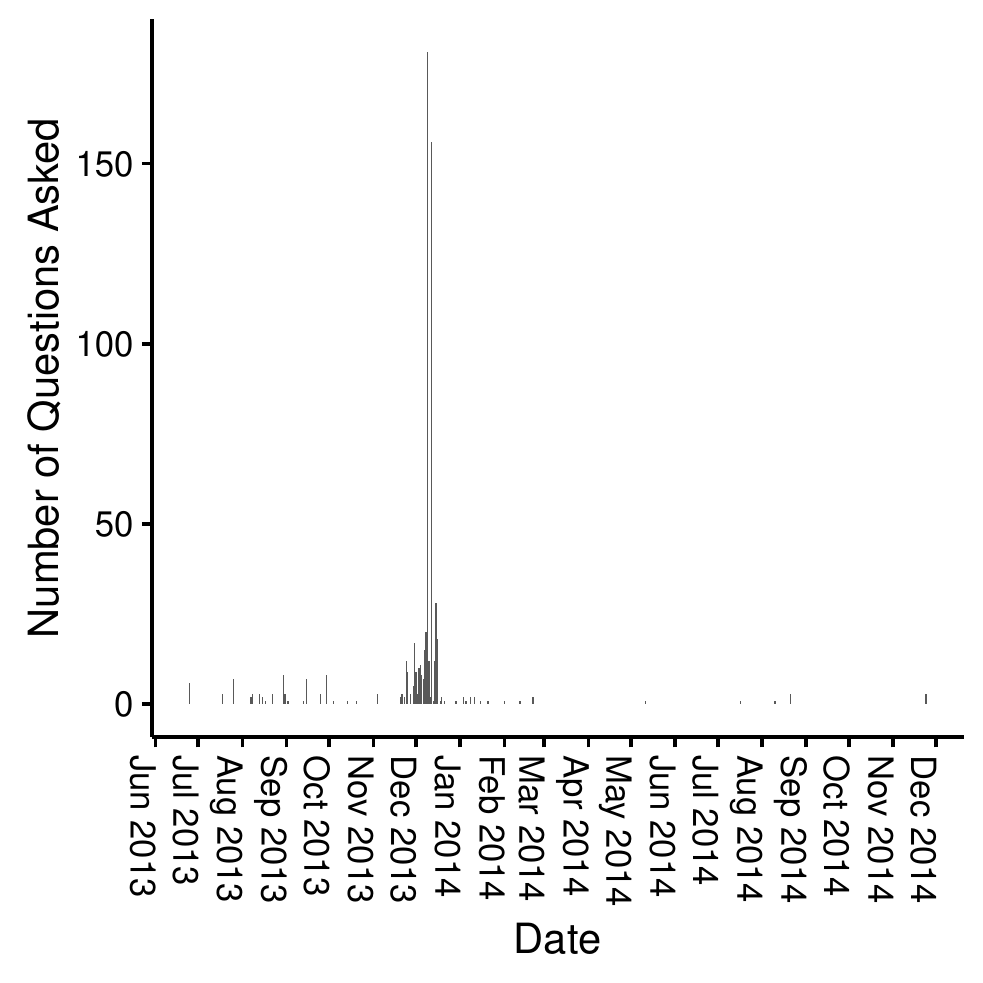}
\caption{Histogram showing the number of questions asked over the study's active time period. Questions were mainly asked in the December, January time period in the winter of 2013/2014. }
\label{fig:question_counts}
\end{figure}

Figure~\ref{fig:question_counts} indicates the number of questions that were asked over time. The vast majority of questions were asked in a short period of time. This period corresponded to a call for participation to BED customers that were received via mail. Motivating continuous participation in such crowdsourced studies versus only initial interest that rapidly wanes would be a valuable problem to endeavor in future work.

\begin{figure}[!ht]
\centering
\includegraphics[width=7cm,height=12cm]{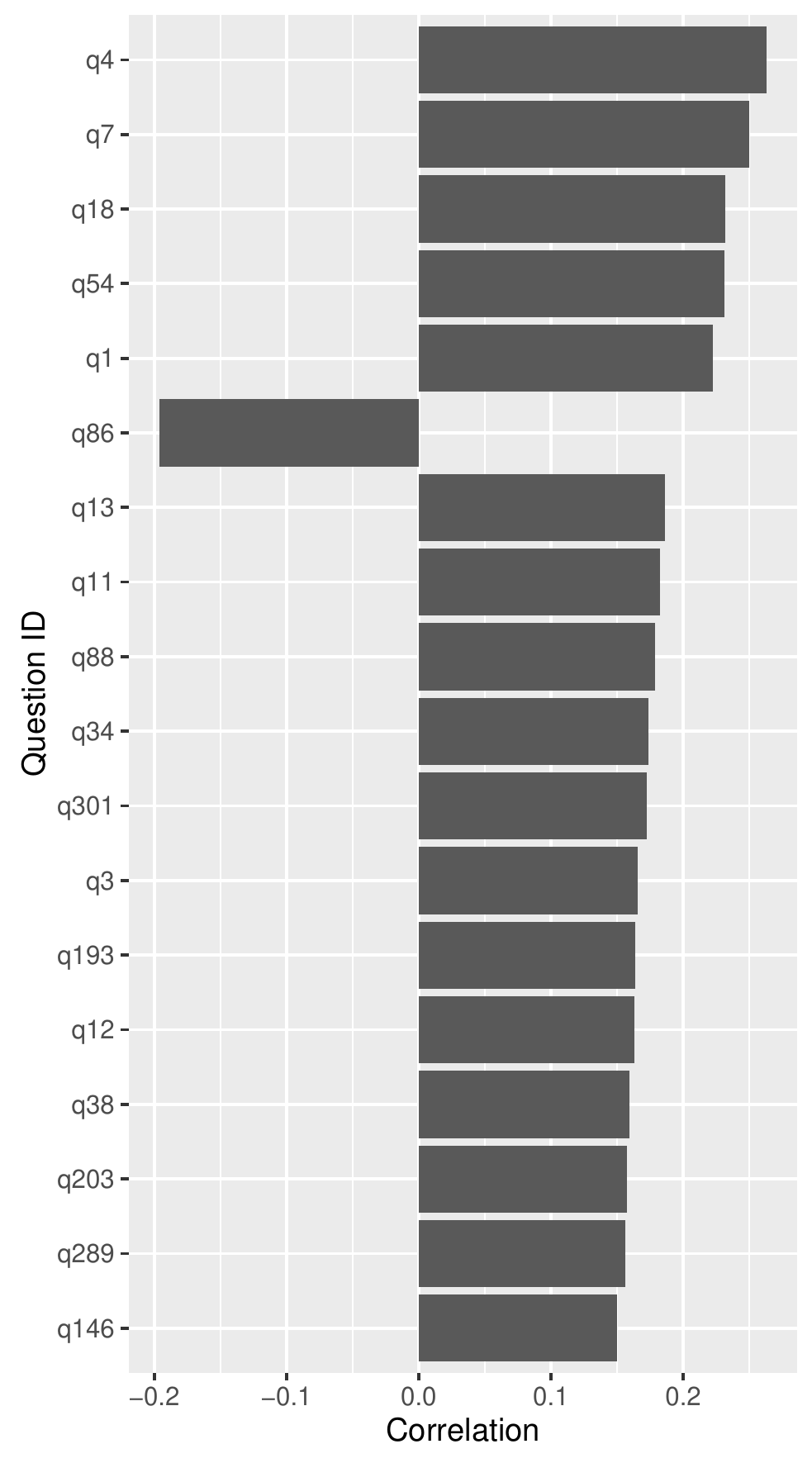}
\caption{Correlations of questions with a correlation value greater than $0.15$. Questions are ranked from highest absolute correlation value to lowest (top to bottom).}
\label{fig:cors}
\end{figure}

Figure~\ref{fig:cors} shows a list of questions whose correlations with the outcome -- residential energy consumed -- were greater than $0.15$. The highest correlated question, \textit{q4} ("How many loads of laundry do you do per week?") was one of the expert-seeded questions seen in Table~\ref{tab:expert_questions}. But the second and third highest correlated questions with the outcome -- \textit{q7} ("How many people live in your household?") and \textit{q18} ("Do you have a dehumidifier?") -- were asked by the 'crowd'.

Note that in the $18$ questions with correlations greater than $0.15$ to the outcome, only one question is negatively correlated. Out of all questions with a correlation of at least $0.01$, only $16\%$ were negatively correlated. That participants appear to focus on questions that are positively correlated with the outcome may be the result of priming, or it could be evidence of a cognitive bias.

\subsection{Predictive Model}

Our true model ($\mathcal{P}$) to null model comparison ($\mathcal{P}_{null}$) does result in a significant difference between the error the predictive model trained on the true dataset, $\mathcal{D}$. Therefore, the models trained using the random forest regression algorithm were able to find some 'signal in the noise'. And thus the questions did indeed provide some degree of predictive power in building the models.

It may be that only the expert provided questions ($q1$ - $q6$) contributed to the predictive ability of $\mathcal{P}$. If this were the case, we would not be able to say that crowd-generated features contributed to a predictive model.

\begin{figure}[!ht]
\centering
\includegraphics[width=0.5\textwidth]{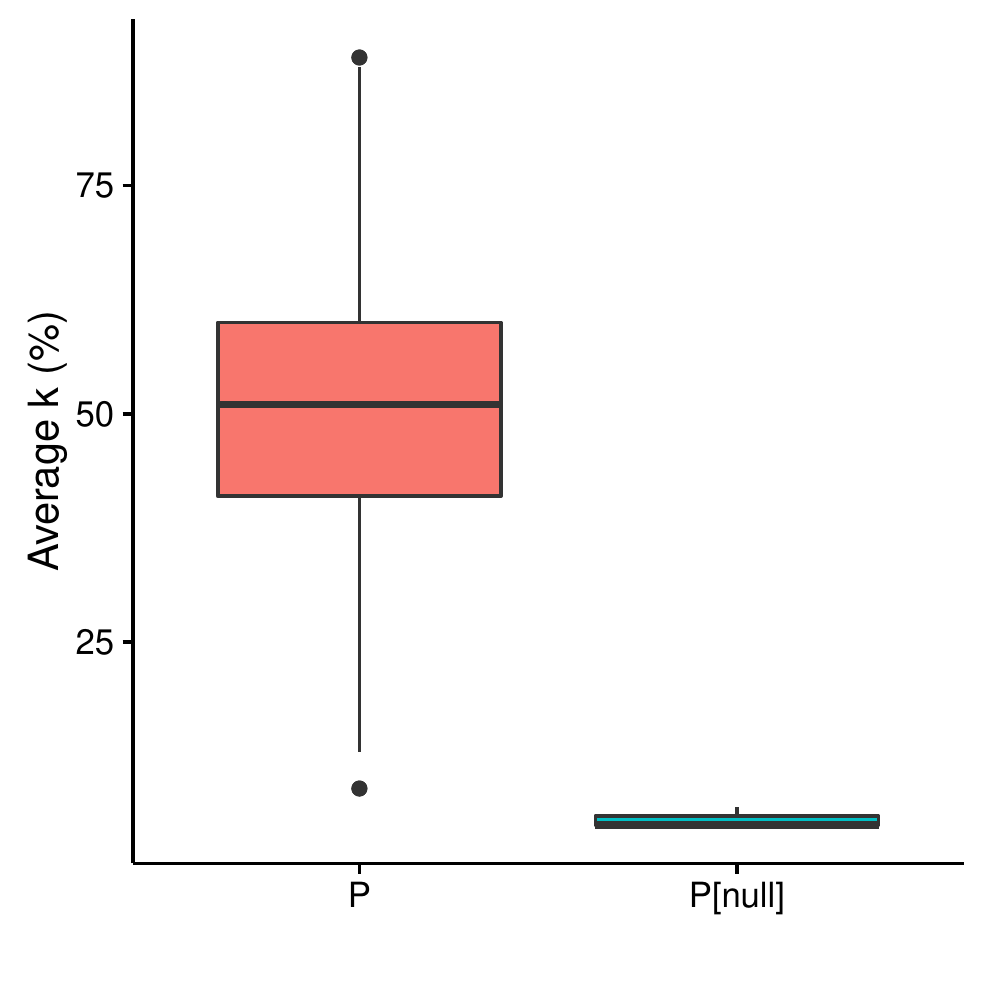}
\caption{Number of top questions used in building $\mathcal{P}$ versus the null model $\mathcal{P}_{null}$ over all parameterizations of $\nu$ and $\delta$. The mode value for the number of top ranked questions in $\mathcal{P}$ was $43$ (median value of $51$), whereas most of the attempts at finding a common cutoff for degeneration, $k$, failed in the question rankings derived from $\mathcal{P}_{null}$.}
\label{fig:avek}
\end{figure}

However, the analysis of the random degeneration of lists indicates that a large number of features used between models are not included by chance alone (on average, the top $51$ ranked questions from $\mathcal{P}$ were considered not to be random between models, see Figure~\ref{fig:avek}). This is in contrast to the null models generated, in which the majority of parameterizations resulted in no valid point at which the feature rankings deviated from meaningful to random collections.

The questions in the list of top ranked questions (Table~\ref{tab:notshuffled}) can be broadly classified into addressing energy choices, lifestyle and behavior (e.g., \textit{q1, q4, q24, q77}); appliances and house features or layout (e.g., \textit{q18, q13, q62, q167}); and house inhabitants (e.g., \textit{q3, q7}). Thus not only did the crowd provide useful predictive features that relate to household appliances and features of the residence that may drive consumption (e.g. size and layout) as are commonly asked in consumer surveys, but they also contributed features related to lifestyle and behavior that are not commonly included in more traditional user surveys. It is in this way that the information resulting from the methodology described here differs from that by utilized by the power utility industry. Appliance saturation surveys, such as the Residential Energy Consumption Survey (RECS) obtain information on number of residents, home features (e.g. age, materials used in construction, size) and information on household appliances. However, in our study consumers were able to provide novel potential drivers of energy consumption that relate to behavior (e.g., ``I only run the dishwasher when it's full'') as well as social factors related to residents (``I know most of my neighbors on a first name basis''). In this way, the crowd has provided novel intuition for what features might impact predictive models that had not previously been considered in traditional industry practices.

Two of the top $10$ questions were expert-generated questions. The probability of two expert questions being found by chance in the top ten is $ \binom{6}{2} \times \binom{626}{8} \div \binom{632}{10} \approx 0.0008$. Thus we have reasonably high confidence that the model incorporates questions that are useful in its top-ranked questions if we are to assume that the expert is able to formulate predictive questions.

Some of the ways that questions contributed cannot be explicitly measured by the models that we have built. For example, whether an earlier user-generated question had an impact on subsequent, possibly more predictive questions, is not readily apparent. For example, a member of the crowd asked the question \textit{q64}, ``Do you generally watch TV in bed at night?''. This may have prompted another user to ask question \textit{q76}, ``How many TVs are in your home?'', which was a high-ranked question in the random forest model. Similarly, the expert-contributed question \textit{q1}, ``I generally use air conditioners on hot summer days'' may have inspired the non-expert user-generated questions \textit{q11} and \textit{q12}, ``How many room/window air conditioners are in your home?'' and ``Do you have central air conditioning?'', respectively. The effect of social influence when asking questions and answering them may have had a positive effect on the overall ability to find predictive behaviors or it could have just as easily negatively influenced the overall crowd effort to explain energy usage. For example, social influence could have caused members of the crowd to become trapped in group pathologies such as groupthink~\cite{janis1972victims}. Details on how the crowd mutually influenced each other is out of the scope of this work, but would be one direction to explore in future work.

Note that some of the predictive questions relate energy consumption to air conditioner use, yet the energy usage data that we are using for training the models is based on data from the winter months. This may at first appear counter-intuitive. It is possible that questions like this uncover general trends in behavior. For example, someone with a low tolerance for discomfort in the summer months also has a low tolerance for discomfort in the winter months. Thus, being the type of person who uses electricity for air conditioners during the summer could be indicative of the same tendency to use heaters that are energy sinks in the winter.

Most of the highest ranked questions appear to related directly to energy usage or behaviors that might clearly affect household energy consumption. However, the seventh-highest-ranked question (\textit{q335}) appears unrelated to energy consumption: ``I know most of my neighbors on a first name basis.'' We can only speculate as to why this question might be predictive of the outcome -- or why the participant who asked the question believed that it would be predictive. That this question was asked and found to be predictive indicates that there may be evidence of a relationship between a person's connection with their neighbors and their energy usage. It is interesting to note that that participants were exposed a graph indicating their own energy usage compared with other participants in the Energy Minder interface. The question may have been influenced by this feature of the website.

\begin{figure}[!ht]
\centering
\includegraphics[width=0.5\textwidth]{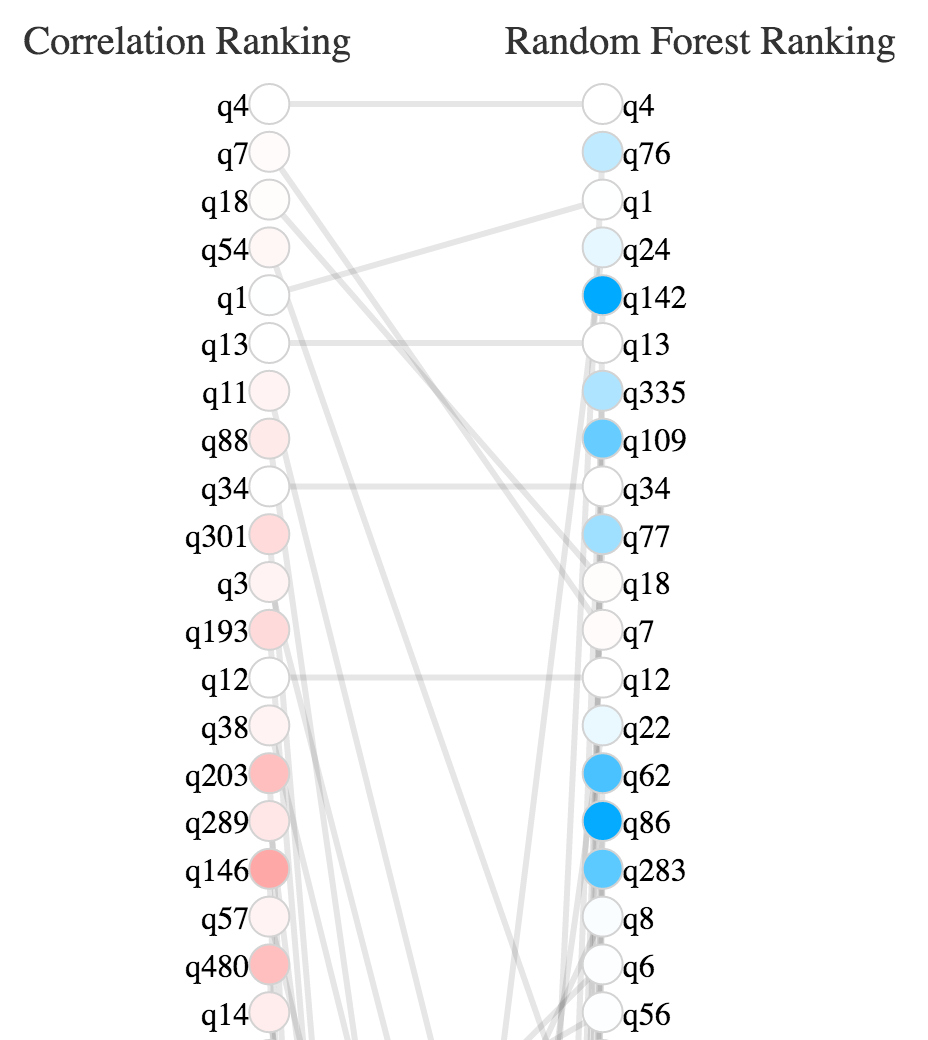}
\caption{Rankings of (top $20$) user-contributed questions by correlation with energy usage value (left column) compared to the rankings of questions computed via the mean Gini impurity ranking method used for importance ranking in the random forest regression algorithm (right column). Top ranked question is at the top (\textit{q4}), the second ranked question by each method is the second from the top (\textit{q7} for correlation and \textit{q76} for random forest regression), and so on. Lines connect questions between rankings. The amount of fill indicates the amount of difference in ranking from one column to another. Red fill in a circle associated with a question indicates a decrease in ranking from left to right and blue for an increase in ranking from left to right. The amount of white in a circle's fill indicates a neutral change in ranking.}
\label{fig:ranks}
\end{figure}

Many of the questions ranked highly by random forests were also highly correlated with the outcome ( Figure~\ref{fig:ranks}). For example, \textit{q4} (``How many loads of laundry do you do per week?'') was most correlated to energy usage and was ranked highest by the random forest algorithm's mean Gini decrease importance criterion. Also, \textit{q1} (``I generally use air conditioners on hot summer days''), \textit{q13} (``How large is your home in square feet of living space?''), and \textit{34} (``How many months of the year do you use your dehumidifier?'') all appear within the top $10$ ranked questions both measured with respect to correlation to the outcome and importance calculated from random forest regression. But there were some instances of questions that were deemed more important by random forest regression than in direct correlation with the outcome. In particular \textit{q142} is ranked $5^{th}$ by random forest regression importance measures and is ranked near last ($595^{th}$) by its correlation to the outcome. Random forests are non-linear regression algorithms. Thus there may be non-linear relationships between such questions as \textit{q142} and energy usage that would not be discovered solely through the use of linear models.

\section{Conclusion and Future Work}
\label{sec:conclusion}
The core contribution of this work was to demonstrate and evaluate a system that enables smart-meter enabled residential electric energy consumers to participate in defining the features of a predictive model.
Consumers were asked to contribute questions that they believed would be predictive of energy usage. They also answered questions about their own usage; and in this way provided data on which to train a regression model. We used a simple regression model to provide immediate feedback to participants on features or questions previously contributed within a web interface. Then we used a \emph{ex post} predictive model using random forest regression to relate total consumer electric energy usage to the user-generated features. In this system, we seek to minimize the error of the predictive model but we do not claim that the modeling methods used -- step-wise AIC linear regression for the web-based tool and random forest regression for the ex post analysis -- are the only or even the best modeling methods for this type of system. In future work, we will investigate whether the use of other modeling methods can in fact improve the overall predictive performance of the system and how the system and crowd are impacted by varying choices.

Our approach uses random forest regression to relate answers to questions posed by the crowd participants themselves, to predict and identify key factors that influence residential electric energy usage. Because of the crowd-driven data collection process, the data were necessarily sparse. This is due to the nature of survey participants both generating the features that the models fit by contributing questions to a survey and by contributing data to these questions by answering them. The sparsity in the dataset is nonuniform due to increasing numbers of questions becoming available as the result of user participation, which then go unanswered by previous participants.

The predictive model resulted in significantly less error than a model trained on a randomized version of the same data. This indicates that the predictive model is indeed finding behavioral drivers of energy consumption. Additionally, we determined where the questions used by the model deviated into only randomly selected features. Questions incorporated into the model up to this cutoff point were thus meaningful contributors to the model's predictive ability. Furthermore, many of the questions that were meaningfully utilized by the predictive model were contributed by non-expert participants. Therefore the crowd contributed in a significant way to the models. The questions incorporated into this models give us clues as to what types of behavior are important to residential energy usage.

While some of these contributed questions were generated by expert users (though also members of the population of residential energy users), the majority of them were produced by non-expert members of `the crowd'. Thus the intuition of the crowd can be used to develop models for residential energy consumption, and possibly other domains.

The methodology outlined here leads to many avenues for future work, including the incorporation of crowd-generated features as complementary to those used in traditional predictive models and the exploration of other machine learning frameworks for achieving better predictability, such as \cite{quadrianto2015very} or \cite{kapelner2015prediction}. Future work should also address whether increasing the number of questions and the number of users improves regression fit and classification accuracy even though the proportion of sparsity stays the same, i.e., whether a 'longer' and 'wider' dataset, but with the same amount of sparsity (i.e., an even larger study) proves beneficial or detrimental to the predictive model.

Additionally, there are opportunities to pose user-generated questions in an adaptive way. For example, it may be possible to motivate the crowd to pose questions that are most likely to attract answers and thus even more explanatory new questions.

There are a number of ways that this collaborative crowdsourcing method can be leveraged to improve planning and operations for electric utilities with Advanced Metering Infrastructure. First, this approach could be a useful component of energy efficiency planning, by identifying large contributors to energy usage in a particular region, without the need for expensive customer surveys. Second, this type of modeling could be extended to identify factors that improve load forecasting. For example, combining this type of data with weather data may identify that customers with certain characteristics are more sensitive to weather changes, enabling more accurate forecasts of load changes in particular portions of a network.

\begin{IEEEbiographynophoto}{Mark D. Wagy} 
received a Ph.D. in Computer Science from the University of Vermont, a B.S. in Computer Science from the University of Minnesota in 2012 and B.A. in Physics and Mathematics from Lewis and Clark College in 2000.

He is currently a post-doctoral research associate in the Department of Psychological and Brain Sciences at Dartmouth College.

\end{IEEEbiographynophoto}

\begin{IEEEbiographynophoto}{Josh C. Bongard} received a Ph.D.~in Computer Science from the University of Zurich in 2003, an M.S. from the University of Sussex in 1999 and a B.S. in Computer Science from McMaster University in 1997.

He is currently the Cyril G. Veinott Professor in the Computer Science Department at the University of Vermont. 
\end{IEEEbiographynophoto}

\begin{IEEEbiographynophoto}{James P. Bagrow} 
received a Ph.D.\ in Physics from Clarkson University in 2008, an M.S.\ (2006) and B.S.\ (2004) in Physics from Clarkson, and an A.S.\ in Liberal Arts and Sciences from SUNY Cobleskill in 2001.

He is currently an Assistant Professor in the Mathematics \& Statistics Department at the University of Vermont.
\end{IEEEbiographynophoto}

\begin{IEEEbiographynophoto}{Paul D.H.~Hines} (M`07,SM`14) received the Ph.D.~in Engineering and Public Policy from Carnegie Mellon University in 2007 and M.S.~(2001) and B.S.~(1997) degrees in Electrical Engineering from the University of Washington and Seattle Pacific University, respectively.

He is currently an Associate Professor, and the L. Richard Fisher professor, in the Dept.~of Electrical and Biomedical Engineering at the University of Vermont. 
\end{IEEEbiographynophoto}

\vfill

\end{document}